\documentstyle[12pt]{ioplppt}
\begin{document}
\jl{6}

\title{Exact nonequilibrium solutions of the Einstein--Boltzmann 
equations. II}

\author{F.P. Wolvaardt 
\dag\ftnote{3}{wolvaardt.f.p@mts400.pgh.wec.com}
and 
Roy Maartens \dag\ddag\ftnote{4}{maartens@sms.port.ac.uk}}
 
\address{\dag\ School of Mathematical Studies, Portsmouth University, 
Portsmouth PO1 2EG, Britain}

\address{\ddag\ Department of Mathematics, 
University of Natal, Durban 4001, South Africa}

\begin{abstract} 

We find exact solutions of the Einstein--Boltzmann 
equations with relaxational
collision term in FRW and Bianchi I spacetimes. The kinematic and 
thermodynamic properties of
the solutions are investigated. 
We give an exact expression for the bulk viscous pressure of an FRW
distribution that relaxes towards collision--dominated equilibrium.
If the relaxation is toward collision--free equilibrium, the bulk
viscosity vanishes -- but there is still entropy production.
The Bianchi I solutions have zero heat flux and bulk viscosity,  
but nonzero shear viscosity.
The solutions are used to construct a realisation of the Weyl
Curvature Hypothesis.

\end{abstract}

\pacs{0420J, 9880H}

\maketitle

\section{Introduction}

In Paper I \cite{mw}, we derived exact properties of the 
Einstein--Boltzmann equations with a
relaxation--time model of collisions.  An exact truncated distribution 
was used to derive transport
equations similar in form to the Israel-Stewart thermodynamics.  
We also found an exact
truncated solution for massless particles in a flat FRW spacetime.  
This paper extends Paper I by
considering exact non--truncated Einstein--Boltzmann solutions 
in FRW and Bianchi I spacetimes.

In Section 2 we present exact anisotropic solutions for a flat FRW 
spacetime,  and investigate
their properties.  In particular we show that the bulk viscosity of 
these solutions is always zero, since they are relaxing towards
free--streaming equilibrium. For solutions that relax toward 
hydrodynamic behaviour, we give an exact formula for the bulk
viscous pressure.
Our solutions generalise the equilibrium anisotropic solutions given 
by \cite{emta}.  Isotropic
non--equilibrium solutions are only possible for massless particles,  
and they have the surprising
property that entropy is generated despite the vanishing of 
bulk viscosity, heat flow, shear viscosity --  and of all non-scalar
moments of the distribution.  This illustrates the point that the 
standard dissipative quantities
(bulk viscosity,  heat flux,  shear viscosity) cannot provide a 
complete or exact description of
non--equilibrium states. 

In Section 3 we find exact anisotropic solutions in Bianchi I 
spacetimes.  The solutions have zero
heat flux and bulk viscosity but nonzero shear viscosity.  
In Section 4 we use the results of the
previous sections to construct a model in which  an 
Einstein--Boltzmann solution in FRW
spacetime evolves into 
an Einstein--Boltzmann solution in Bianchi spacetime.  The basic 
idea,  due to Matravers and Ellis
\cite{me},  is that anisotropy in the FRW distribution is 
communicated to the geometry via the
Einstein field equations when the gas cools sufficiently for 
massive particles to become
non--relativistic.  We provide an explicit dissipative realisation of 
the Ellis--Matravers model,  which
is in accord with Penrose's Weyl Curvature Hypothesis,  i.e.  
that the universe is initially
conformally flat and that anisotropy,  inhomogeneity and 
entropy production develop as the
universe expands \cite{p1},  \cite{p2}. 

The distribution $f(x,p)$ satisfies the Boltzmann equation with 
relaxational collision term,  i.e.  the
BGK equation \cite{mw}
\begin{equation}
L[f] \equiv {df \over dv} \equiv  p^i {\partial f \over \partial x^i} 
- \Gamma^i{}_{jk} p^j p^k {\partial f
\over \partial p^i} = \gamma (x,E) ( \bar{f} - f )
\label{bgk}
\end{equation}
where $\gamma$ encodes microscopic interaction information in a 
linear, macroscopic approximation \cite{mw},
$\bar{f}$ is the distribution towards which (or
away from which if $\gamma < 0$) $f$ is relaxing, and $E=-u_ip^i$ is 
the particle energy relative to the average four--velocity $u^i$
associated with $\bar{f}$.
If $\bar{f}$ is a dynamic (or `global') equilibrium 
distribution \cite{mw},  $L [ \bar{f}] = 0$,  then the solution of 
(\ref{bgk}) is 
\begin{equation}
f = \bar{f} + h e^{- \Gamma} \mbox{  where  }
L[h]=0\,,\quad \Gamma\left(x(v),E(v)\right) = \int^v 
\gamma\left(x(u),E(u)\right) du
\label{solf}
\end{equation}
The standard form for $\gamma$ is the Anderson--Witting (AW) 
form \cite{aw}:
\begin{equation}
\gamma (x,E) = {E \over \tau (x)}
\label{aw}
\end{equation}
where $\tau$ is the mean interaction time. This form includes the
case of radiative transfer with isotropic scattering \cite{ui}.
Using $dv=dt/p^0=dt/E$ and (\ref{aw}), the relaxation factor
in (\ref{solf}) becomes a spacetime scalar:
\begin{equation}
\Gamma=\int{dt\over\tau}
\label{G}\end{equation}
The distribution
may be covariantly decomposed relative to $u^i$ \cite{emtb}:
\begin{equation}
f(x,p) = F(x,E) + F_i (x,E)e^i +  F_{ij} (x,E)e^i e^j + \cdots
\label{covh}
\end{equation}
where the covariant multipoles $F_{ij \cdots}$ are isotropic,  
spatial,  trace-free and symmetric, 
and $e^i$ is the unit spatial projection of $p^i$. 

In order to avoid unnecessary details,  we will not summarise the 
basic equations and results of
Paper I,  but refer where necessary directly to the equations 
in that paper in the form (I:$n$), where
$n$ is the equation number in Paper I. 

\section{Exact non--equilibrium Einstein--BGK solution in FRW 
spacetime }

In Paper I we found an Einstein--BGK solution for a truncated form 
of the distribution function. 
This solution was used to derive a set of exact thermodynamic laws.  
In this section we present
the full non--truncated solution in flat FRW spacetime with 
natural coordinates $x^i = (t,x^\nu)$:  
\begin{equation}
ds^2 = -d t^2 +R^2 (t)(dx^2 +dy^2+dz^2)
\label{rwm}
\end{equation}
By homogeneity,  the spatial momenta $p_\nu$ are constants of the 
motion,  i.e.  $L[p_\nu]=0$. 
Thus homogeneous distributions are of the form $f(t,p_\nu)$,  
while isotropic and homogeneous
distributions are of the form $f(t,w)$ where
\begin{equation}
w^2 \equiv (p_1)^2 + (p_2)^2 + (p_3)^2 = (E^2 - m^2) R^2
\label{3mom}
\end{equation}
and where $u^i=\delta^i{}_0$.
If the homogeneous $f$ is relaxing toward the homogeneous and
isotropic equilibrium,  $\bar{f}$,  then by
(\ref{solf}) and (\ref{G})
\begin{equation}
f(t,p_\nu) = \bar{f} (w) + h(p_\nu)e^{-\Gamma(t)}\,,\quad
\Gamma(t)=\int_0^t{dt'\over\tau(t')}
\label{rwf}
\end{equation}
Clearly $f$ is spatially homogeneous but dynamically anisotropic, and 
it must depend explicitly on
the cosmic time (i.e. $\dot{\Gamma}\neq0$)
if it is non--equilibrium.  
Note that since $L[\bar{f}]=0$, $\bar{f}$ can only be a
collision--dominated Maxwell--Boltzmann equilibrium if $m=0$;
for $m > 0$,  $\bar{f}$
must be a collision--free equilibrium distribution \cite{mw}.  
In this case the BGK solution $f$ represents a
distribution relaxing towards free--streaming isotropic equilibrium. 

In order to investigate the properties of the solutions, we use the 
covariant harmonic
decomposition (\ref{covh}):
\begin{equation}
F(t,w) = \bar{F} (w) + H(w) e^{- \Gamma (t)}\,,\quad
F_{\nu \cdots \kappa}(t,w) = H_{\nu\cdots\kappa}(w) e^{-\Gamma (t)}
\label{rwcovh}
\end{equation}

The kinematics and the dynamics of the solution (\ref{rwcovh}) are 
determined by the particle
four--current $N^i = n u^i + k^i$ and energy--momentum tensor 
$T_{ij} = \mu u_i u_j + ph_{ij} +
\pi_{ij} + q_i u_j +q_j u_i$, where $n$ is the number density,  $k^i$ 
is the number flux,  $\mu$ is
the energy density,  $p$ is the isotropic pressure, $h_{ij}=g_{ij}+
u_i u_j$ is the spatial projector,
$\pi_{ij}$ is the anisotropic pressure tensor and
$q_i$ is the heat flux.  These quantities follow from (I:40)
and (\ref{rwcovh}):
\begin{eqnarray}
 n&=&\bar{n}+{4\pi e^{-\Gamma(t) }\over{R^3(t)}}
 \int _0^\infty w^2H(w)dw \label{prop1}\\
k_\nu&=&{4\pi e^{-\Gamma(t) }\over{3R^3(t)}}\int _0^\infty 
w^3[w^2+m^2R^2(t)]^{-1/2}H_\nu (w)dw  \label{prop2}\\
\mu&=&\bar{\mu}+{4\pi e^{-\Gamma(t)}\over{R^4(t)}}\int ^\infty _0
w^2[w^2+m^2R^2(t)]^{1/2}H(w)dw \label{prop3}\\
p&=&\bar{p}+{4\pi e^{-\Gamma(t)}\over{3R^4(t)}}\int ^\infty _0
w^4[w^2+m^2R^2(t)]^{-1/2}H(w)dw \label{prop4}\\ 
q_\nu&=&{4\pi e^{-\Gamma(t)}\over{3R^4(t)}}\int _0^\infty 
w^3H_\nu (w)dw \label{prop5}\\
\pi _{\nu\kappa}&=&{8\pi e^{-\Gamma(t)}\over{15R^4(t)}}\int _0^\infty 
w^4[w^2+m^2R^2(t)]^{-1/2}H_{\nu\kappa}(w)dw  \label{prop6}
\end{eqnarray} 
with $k_0=0$, $q_0=0$ and $\pi _{0i}=0$.  
A non--zero number flux gives a particle drift that is out
of keeping with FRW symmetry, and although it is possible to satisfy 
the field equations for $k_i\neq 0$ (see the fluid solutions 
of Calvao and Salim \cite{cs}), we regard this as unnatural. 
From equation (\ref{prop2}) it is clear that in order to get a zero 
number flux,  which gives a
non--tilted kinematic average 4--velocity,  $H_\nu$ must vanish 
for $m>0$.  It is possible to find
nonzero $H_\nu$ if $m=0$:
\begin{equation}
\int_0^\infty w^2H_\nu(w)dw = 0
\label{nflux0}
\end{equation}
The full Boltzmann collision term is based on microscopic conservation 
so that the macroscopic
conservation of momentum and energy are identically satisfied.  This 
is not the case for the BGK
collision model, and the conditions imposed by the conservation 
equations require separate
investigation \cite{mw}. The conservation of particle number, energy 
and momentum are given by
equation (I:48).   On using the FRW BGK solution (\ref{rwcovh}),  
we find the following
condition for the conservation of particles: 
\begin{equation}
\int _0^\infty w^2H(w)dw=0 
\label{pcon}
\end{equation}
which by (\ref{prop1}) implies $n=\bar{n}$,
i.e. the number density is matched to that of the limiting
equilibrium.  

The condition for the conservation of energy gives
\begin{equation}
\int _0^\infty w^2[w^2+m^2R^2(t)]^{1/2}H(w)dw=0
\label{econ}
\end{equation}
If $m>0$, this forces $H=0$, but $H$ may be nonzero for $m=0$.  
In all cases, (\ref{econ}) in (\ref{prop3}) implies
$\mu = \bar{\mu}$,  so that the energy density matches the
equilibrium value.

Momentum conservation gives
\begin{equation}
\int _0^\infty w^3H_\nu(w)dw=0
\label{mcon}
\end{equation}
Condition (\ref{mcon}) is automatically satisfied for $m>0$, since
$H_\nu=0$ from $k_\nu=0$. For $m=0$, it is a further condition.
By (\ref{prop5}), we see that (\ref{mcon})
leads to $q_\nu = \bar {q}_\nu = 0$,  so that $u^i$ is also the
energy--frame four--velocity (it is already the particle--frame
four--velocity by $k_\nu=0$, so the two four--velocities coincide
in these solutions).  

The entropy density $s$ is given by (I:40g).  
For the FRW solution (\ref{rwcovh}):
\begin{eqnarray}
s&=&{{4\pi }\over{R^3(t)}}\int ^\infty _0w^2\left\{\left[\bar{F}(w)
+H(w)e^{-\Gamma(t)}\right]\left[1-\ln[\bar{F}(w)
+H(w)e^{-\Gamma(t)}]\right]  \right. \nonumber \\
& & {}- \left.
{\textstyle{1\over6}}\left[\bar{F}(w)+H(w)e^{-\Gamma(t)}\right]^{-1}
e^{-2\Gamma(t)}
H_\nu(w)H^\nu(w) + \cdots \right\} dw
\label{s}
\end{eqnarray}
The entropy production rate (I:40i) becomes
\begin{eqnarray}
S^i{}_{;i}&=&{4\pi e^{-\Gamma(t)}\over{R^3(t)\tau(t)}}
\int ^\infty _0 w^2
\left\{ H(w) \ln \left[ \bar{F}(w)+H(w)e^{-\Gamma (t)} \right]
\right. \nonumber \\
& &{}+{\textstyle{1\over6}}e^{-\Gamma(t)}H_\nu(w) H^\nu(w) 
\left[ \bar{F}(w) +
H(w)e^{-\Gamma(t)} \right]^{-2} \times\nonumber\\ 
&&{}\times\left.\left[ 2\bar{F}(w)+H(w)
e^{-\Gamma (t)}\right]+\cdots \right\} dw
\label{erate}
\end{eqnarray}
For the Boltzmann collision term, $S^i{}_{;i}\geq0$ follows 
identically,  but this is not true for the
BGK collision term \cite{mw}. It is not clear in general 
from (\ref{erate}) whether the H--theorem,
$S^i{}_{;i}\geq0$,  is satisfied without restrictions for the BGK 
solution (\ref{solf}).  This needs to
be checked for each specific solution.  Note that for $m>0$,  
when $H=0=H_\nu$,  the only
contribution to $S^i{}_{;i}$ is from the quadrupole and higher 
moments.  For $m=0$,
$H$ is in general nonzero and there is a monopole contribution 
to the entropy production:
\begin{equation}
S^i{}_{;i}={{4\pi e^{-\Gamma(t)}}\over{R^3(t)\tau(t)}}
\int^\infty_0 w^2H(w) \ln [\bar{F}(w) + H(w)
e^{-\Gamma(t)}]dw + \cdots
\label{eaw}
\end{equation}

In summary:
\begin{eqnarray}
 H=0=H_\nu &~\mbox{ for }~& m>0 \label{aw1}\\
\int _0^\infty w^rH(w)dw=0=\int _0^\infty w^rH_\nu(w)dw
&~\mbox{ for }~& m=0 \quad (r=2,3) \label{aw2}
\end{eqnarray}
With
(\ref{aw1}) and  (\ref{aw2}),  conservation of particle number and 
energy--momentum is satisfied and
we have 
\begin{equation}
n=\bar{n}\,,~ \mu=\bar{\mu}\,,~ k_\nu=0\,,~ q_\nu=0
\label{aw3}
\end{equation}\ 
Now we impose the Einstein field equations (I:61) 
(note that the conservation equations are
already satisfied):
\begin{eqnarray}
q_\nu&=&0=\pi_{\nu\kappa}
\label{ef1}\\
\mu&= &3{\dot{R}^2 \over R^2}
\label{ef2}
\end{eqnarray}
The heat flux is already zero by (\ref{aw3}).
Vanishing anisotropic stress requires,  by (\ref{prop6})
\begin{equation}
\int _0^\infty w^4[w^2+m^2R^2(t)]^{-1/2}H_{\nu\kappa}(w)dw=0
\label{0stress}
\end{equation}
For $m>0$ (\ref{0stress}) forces $H_{\nu\kappa}=0$.  For $m=0$,  
$H_{\nu\kappa}$ is subject to
\begin{equation}
\int _0^\infty w^3H_{\nu\kappa} (w)dw=0
\label{stressaw}
\end{equation}
Using (\ref{prop3}) in the Friedmann equation (\ref{ef2}) 
we can write it as
\begin{eqnarray}
 \dot{R}(t)&=&\left[ {{4\pi }\over{3R^2(t)}}\int 
^\infty_0w^2[w^2+m^2R^2(t)]^{1/2}\bar{F}(w)dw \right. \nonumber \\
& &{}+ \left. {{4\pi e^{- \Gamma(t)} \over{3R^2(t)}}\int^\infty _0w^2
[w^2+m^2R^2(t)]^{1/2}H(w)}dw \right] ^{1/2}
\label{rdot}
\end{eqnarray}
By (\ref{aw1}) and (\ref{aw2}) the term containing $H(w)$ 
in (\ref{rdot}) vanishes for $m\geq 0$
and we can re--arrange this equation and give the solution explicitly 
as 
\begin{equation}
t={{\sqrt{3}}\over{2}}\int_0^{R^2}\left[4\pi
\int^\infty_0w^2[w^2+m^2u]^{1/2}\bar{F}(w)dw\right]^{-1/2}du
\label{cosmict}
\end{equation}
which is the same as for an equilibrium Einstein solution \cite{emta}.  
By specifying $\bar{F}$, 
$R(t)$ can be determined in principle,  and the metric (\ref{rwm}) 
will be known -- completing the
Einstein solution.  
This is analogous to specifying an equation of state in 
a fluid model. 

The reason that $R$ has the same form as for an equilibrium 
solution lies in the vanishing of the bulk viscous pressure 
$\Pi = p - \bar{p}$.  By (\ref{prop4})
\begin{equation}
\Pi= {4 \pi e^{-\Gamma} \over 3R^4}\int_0^\infty 
w^4[w^2 + m^2R^2]^{-1/2} H(w)dw
\label{Pi}
\end{equation}
It follows from (\ref{aw1}) and (\ref{aw2}) that $\Pi=0$ for $m\geq0$.  
For $m=0$,  this is in
accord with the approximation schemes used to derive transport 
equations \cite{is}.  For $m>0$, 
the vanishing of bulk viscosity is a consequence of our choice 
of $\bar{f}$ satisfying
$L[\bar{f}]=0$.  This means that $\bar{f}$ cannot be a 
Maxwell--Boltzmann distribution (unless the
universe is static) \cite{mw},  and therefore $f$ cannot be 
approaching the hydrodynamic regime
where the standard approximation schemes are applied and 
predict $\Pi\neq0$.  Instead $f$ must
be collision--free,  and if $f$ is near to free--streaming ,  
our results show that,  unsurprisingly,  
there is no bulk viscosity. 

If we drop the restriction $L[\bar{f}]=0$ and take $\bar{f}$ 
to be a local Maxwell--Boltzmann
distribution,  we have \cite{mw}
\begin{equation}
\bar{f}= \exp{\left[\alpha(t) - {E\over{T(t)}}\right]}
\label{mb}
\end{equation}
where $T$ is the temperature and $\alpha$ the chemical potential.  
Using the general (i.e. $L[\bar{f}]\neq0$) BGK solution
(I:44),  we find the following exact formula for $\Pi$:
\begin{eqnarray}
\Pi&=&-{4\pi \over 3} e^{-\Gamma} \int_0^t dt\, e^{\Gamma+\alpha}
\int_m^\infty dE\, (E^2-m^2)^{3/2}e^{-E/T}\,\times \nonumber\\
&&{}\times\,\left\{ \dot{\alpha} + 
{E\over T}\left[{\dot{T}\over T}+{\dot{R}\over
R}\left(1-{m^2\over E^2}\right)\right]\right\}
\label{Pimb}
\end{eqnarray}
In the case $m=0$, when $T\propto R^{-1}$ and
$\dot{\alpha}=0$, we have $\Pi=0$, as expected.

What is remarkable about the solutions with $L[\bar{f}]=0$
is this:  despite the vanishing of viscosity and heat flux, 
there is dissipation,  since $L[f] \neq 0$.  
The simplest case is the isotropic massless 
solution ($\alpha$ constant)
\begin{equation}
f= \bar{f} + He^{-\Gamma (t)}, \mbox{  } \bar{f} = e^{\alpha - w}
\label{iso}
\end{equation}
first presented in \cite{mw}.  By (\ref{eaw}) and (\ref{aw2}) the 
entropy production rate is 
\begin{equation}
S^i{}_{;i} = {{4 \pi e^{-\Gamma(t)}} \over{R^3(t)\tau(t)}} 
\int^\infty_0
w^2H(w)\ln[1+H(w)e^{w-\Gamma (t)- \alpha}]dw \nonumber
\end{equation} 
and this is clearly positive since $f$ is.  
The point is that the bulk
viscosity,  heat flux and shear viscosity are only approximate 
indicators of non--equilibrium states, 
and even if they vanish there can still be dissipation.  
Non--equilibrium states in general cannot be
completely,  and certainly not exactly,  described by these 
standard dissipative quantities. 
Usually this is understood in terms of the effect of multipoles 
higher than the quadrupole,  which
are neglected in the standard approximation schemes.  However,  the 
isotropic solution (\ref{iso})
has no multipoles beyond the scalar monopole,  and yet it is 
out of equilibrium. 

\section{Exact non--equilibrium Einstein--BGK solutions 
in Bianchi I spacetime}

Bianchi I spacetime,
\begin{equation}
ds^2=-dt^2+X^2(t)dx^2+Y^2(t)dy^2+Z^2(t)dz^2 
\label{b1m}
\end{equation}
is distinguished kinematically from FRW spacetime by 
non--zero shear,  whose evolution is given
by \cite{me}
\begin{equation}
\dot{\sigma}_{ij}-\sigma _{ik}{\sigma ^k}_j
+{\textstyle{1\over3}}h_{ij}
\sigma^{kl}\sigma_{kl}-{\textstyle{1\over2}}\pi _{ij}+E_{ij}=0
\label{eos}
\end{equation}
where $E_{ik}=C_{ijkl}u^ju^l$ is the electrical part of the 
Weyl tensor $C_{ijkl}$. By symmetry, the trace--free spatial
tensors have the form
\begin{equation}
A^i{}_j=\mbox{diag}\left(0,A^1{}_1,A^2{}_2,-A^1{}_1-A^2{}_2\right)
\label{sym}\end{equation}
where $A_{ij}=\sigma_{ij},\pi_{ij},E_{ij}$ or $F_{ij}$.

The Einstein field equations for this metric are \cite{me}
\begin{eqnarray}
{\ddot{X}\over{X}}+{\ddot{Y}\over{Y}}+{\ddot{Z}\over{Z}}
&=&-{\textstyle{1\over2}}(\mu+3p)\label{bief1}\\ 
q_j&=&0\label{bief2}\\
{\ddot{X}\over{X}}+{\dot{Y}\over{Y}}{\dot{X}\over{X}}+
{\dot{Z}\over{Z}}{\dot{X} \over{X}}
&=&{\textstyle{1\over2}}(\mu-p)+\pi^1{}_1\label{bief3}\\
{\ddot{Y}\over{Y}}+{\dot{Y}\over{Y}}{\dot{X}\over{X}}+
{\dot{Z}\over{Z}}{\dot{Y} \over{Y}}
 &=&{\textstyle{1\over2}}(\mu-p)+\pi^2{}_2\label{bief4}\\
{\ddot{Z}\over{Z}}+{\dot{Z}\over{Z}}{\dot{X}\over{X}}+
{\dot{Z}\over{Z}}{\dot{Y} \over{Y}}
 &=&{\textstyle{1\over2}}(\mu-p)-\pi^1{}_1-\pi^2{}_2 \label{bief5}
\end{eqnarray}
The conservation of energy--momentum reduces to
($\Theta=u^i{}_{;i}$)
$$
\dot{\mu}+(\mu +p)\Theta +\pi _{ij}\sigma ^{ij}=0 
$$
which is identically satisfied if (\ref{bief1}) -- (\ref{bief5}) 
are satisfied. 

By homogeneity, the spatial momenta $p_\nu$ are constants 
of the motion.  Thus
homogeneous distributions are of the form $f(t,p_\nu)$,
including the special case $f(t,w)$ where
\begin{eqnarray}
 w^2&=&(p_1)^2+(p_2)^2+(p_3)^2\nonumber \\ 
&=&X^4(t)(p^1)^2+Y^4(t)(p^2)^2+Z^4(t)(p^3)^2\label{b1w}
\end{eqnarray}
Contrary to the FRW case,  $w$ is {\em not} isotropic, i.e. it is
not a function only of $E$ in momentum space, where
\begin{equation}
E=[m^2+X^{-2}(t)(p_1)^2+Y^{-2}(t)(p_2)^2+Z^{-2}(t)(p_3)^2]^{1/2}
\end{equation}
Both $w$ and $E$ are 
spatially homogeneous.  However, there is no simple relation 
between the anisotropic 
constant of motion $w$ and the isotropic non--constant energy $E$,  
unlike the FRW case.  This
happens since $h_{ij}$ is anisotropic:
\begin{eqnarray}
\lambda ^2&\equiv&E^2-m^2=h^{\mu \nu }p_\mu p_\nu \nonumber \\
w^2&=&\delta ^{\mu \nu }p_\mu p_\nu\nonumber 
\end{eqnarray}
In the FRW case $h^{\mu \nu }=R^{-2}(t)\delta ^{\mu \nu }$. 

If the homogeneous,  anisotropic $f$ is relaxing toward the 
homogeneous,  anisotropic $\bar{f}$, where $L[\bar{f}]=0$,
then the BGK solution has the same form (\ref{rwf}) as in the
FRW case, but with $w$ given by (\ref{b1w}) (and anisotropic).  
It follows that the Bianchi
I solution matches the FRW solution (\ref{rwf}) in the limit 
$R(t)=X(t)=Y(t)=Z(t)$.  
The covariant
harmonic decomposition (\ref{covh}) takes the form
\begin{equation}
f(t,p_\nu )=F(t,E)+F_\kappa(t,E)e^\kappa+F_{\kappa\rho}(t,E)
e^\kappa e^\rho + \cdots
\label{bicovh}
\end{equation}
where 
$$
F_{\nu\cdots\kappa}(t,E)=\bar{F}_{\nu\cdots\kappa}(t,E)+
e^{-\Gamma (t)} H_{\nu\cdots\kappa}(t,E)
$$ 
Note that
since $\bar{f}$ is anisotropic,  its higher order multipoles cannot be 
neglected in the decomposition. 
Although the solution and its decomposition do not take the 
convenient form of the FRW case, it 
gives us the necessary tools to investigate the conditions for 
an Einstein--BGK solution. 

Because of the nonvanishing shear in the Bianchi I geometry,  
the covariant multipoles in
(\ref{bicovh}) are no longer independent.  The relationship between 
the multipoles is determined
by the Boltzmann equation.  By attaching an orthonormal tetrad to 
$u^i$, the Liouville operator
$L$ can be written in the covariant harmonic form (I:10).  
This allows one to write the Boltzmann
equation as a set of coupled differential equations in 
the multipoles.  
The first two Boltzmann multipole equations for a
homogeneous distribution function \cite{emtc}
become, for a Bianchi I geometry and a
BGK AW collision term:
\begin{eqnarray}
&& {2\over15}\lambda ^{-1}{\partial\over\partial E}\left(
\lambda ^3\sigma^{\nu\kappa}F_{\nu\kappa}\right) 
+{1\over3}\lambda ^2\Theta{\partial F\over\partial E}\nonumber\\
&&{}- E{\partial F\over\partial t}
=\tau^{-1}E(F-\bar{F})\label{bhe1}\\
&&-{6\over35}\lambda^{-2}{\partial\over\partial E}\left(\lambda^4
\sigma^{\kappa\rho}F_{\nu\kappa\rho}\right)
+{2\over5}\lambda^{1/2}{\partial\over
\partial E}\left(\lambda^{3/2}\sigma_{\nu\kappa}
F^\kappa\right)\nonumber\\
&&{}+{1\over3}\lambda^2\Theta{\partial F_\nu\over\partial E}
=\tau^{-1}E(F_\nu-\bar{F}_\nu) \label{bhe2}
\end{eqnarray}
where we have used the Bianchi I symmetry to simplify the expressions
given in \cite{me}.
The higher order multipole equations \cite{emtc} (p492) 
together with (\ref{bhe1}), (\ref{bhe2}) show that if
$F$, $F_\nu$ are specified, then the multipole equations place a 
chain of restrictions on the
quadrupole and higher moments.  The important point is that if 
the shear  
is non--zero, the multipoles are no longer independent. 

From (I:40),  the kinematic and dynamic quantities of
the solution (\ref{bicovh}) are
\begin{eqnarray}
n&=&\bar{n}+4\pi e^{-\Gamma(t)} \int _m^\infty E\lambda 
H(t,E)dE\label{bprop1}\\ 
k_\nu&=&\bar{k}_\nu+{{4\pi }\over{3}}e^{-\Gamma(t)}\int _m^\infty 
\lambda^2 H_\nu(t,E)dE\label{bprop2}\\ 
\mu& =&\bar{\mu} +4\pi e^{-\Gamma(t)} \int _m^\infty E^2\lambda
H(t,E)dE\label{bprop3}\\
p&=&\bar{p}+{{4\pi }\over{3}}e^{-\Gamma(t)}\int _m^\infty \lambda ^3
H(t,E)dE\label{bprop4}\\
q_\nu&=&\bar{q}_\nu+{{4\pi }\over{3}}e^{-\Gamma(t)}\int _m^\infty
E\lambda^2H_\nu(t,E)dE\label{bprop5}\\ 
\pi _{\nu\kappa}&=&\bar{\pi}_{\nu\kappa}+{{8\pi }\over{15}}
e^{-\Gamma(t)}\int _m^\infty \lambda ^3 H_{\nu\kappa}(t,E)
dE\label{bprop6}
\end{eqnarray}
The particle flux $\bar{k}_i$, energy flux $\bar{q}_i$ and anisotropic 
stress $\bar{\pi}_{ij}$ of the 
equilibrium distribution $\bar{f}$ are in
general nonzero because $\bar{f}$ is anisotropic. These quantities do
not reflect any dissipation, but are part of the measure of 
deviation from isotropy (compare \cite{mes}, \cite{nm}).

Given that the BGK collision model does not guarantee macroscopic 
conservation,  we impose
the conditions for the conservation of particle number, energy and 
momentum (I:40g):
\begin{eqnarray} 
\int _m^\infty E\lambda H dE &=&0 \label{bncon}\\
\int _m^\infty E^2\lambda H dE &=&0 \label{becon}\\
\int _m^\infty E\lambda^2 H_\nu dE &=&0 \label{bmcon}
\end{eqnarray} 
With (\ref{bprop1}),  (\ref{bprop3}) and (\ref{bprop5}) these
give the matching conditions:
\begin{equation}
n=\bar{n},\mbox{  }\mu =\bar{\mu}, \mbox{  }q_\nu=\bar{q}_\nu
\label{bmatch}
\end{equation} 
The Bianchi I solution has equilibrium particle number density, energy
density and energy flux. In general, since $H$ and $H_\nu$ depend
explicitly on time, (\ref{bncon}) -- (\ref{bmcon}) require
\begin{equation}
H=0=H_\nu
\label{bcon1}
\end{equation}

The Einstein field equations are imposed next.  Equation (\ref{bief2}) 
along with (\ref{bprop5}) and (\ref{bmcon}) implies
$$
\int_m^\infty E^2 \lambda^2\bar{F}_\nu dE=0
$$
Again, in general this requires
\begin{equation}
\bar{F}_\nu = 0 
\label{bief1a}\end{equation}
Equations (\ref{sym}) and (\ref{bprop6}) require 
\begin{equation}
\int ^\infty _m\lambda ^3F^\nu{}_\kappa(t,E)dE={{15}\over{8\pi}}
\mbox{diag}\left(\pi^1{}_1,\pi^2{}_2,-\pi^1{}_1-\pi^2{}_2\right)
\label{bief2a} 
\end{equation}
Thus $F_{ij}$ has at most two independent components.  
Then (\ref{bhe1}), (\ref{bhe2}), (\ref{bmatch}), 
(\ref{bief1a}) and (\ref{bief2a}) are the restrictions on the 
harmonics for an Einstein--BGK solution. 
Once $F$ and $F^\nu{}_\nu$ (no sum) are specified,  
the remaining field equations in principle yield a solution
$\{X(t),Y(t),Z(t)\}$ (this is analogous to specifying equations of 
state for $p$ and $\pi _{ij}$ in fluid
models). 

By (\ref{bprop4}) the bulk viscous pressure $\Pi$ is given by
\begin{equation}
\Pi ={4\pi\over3}e^{-\Gamma(t)}\int _m^\infty \lambda ^3H(t,E)dE
\end{equation}
It follows from (\ref{bmatch}) and (\ref{bcon1}) 
that $\Pi = 0$ for $m \geq 0$.\ Again the 
bulk viscosity vanishes when $m>0$ because $f$ is relaxing toward
free--streaming.

The restrictions (\ref{bcon1}), (\ref{bief1a}) 
and (\ref{bief2a}) on the multipoles may for example
be satisfied by the choice:
\begin{equation}
F=\bar{F}\,,~F_\nu=0=\bar{F}_\nu\,,~\bar{F}_{\nu\kappa}=0\,,~
H^\nu{}_\kappa=\mbox{diag}\left(V_1,V_2,-V_1-V_2\right)
\label{b1eg} 
\end{equation}
where
$$
\pi^\nu{}_\nu={8\pi\over15}e^{-\Gamma(t)}\int_m^\infty \lambda^3
V_\nu(t,E)dE\quad\mbox{(no sum)}
$$
This choice implies that the equilibrium distribution $\bar{f}$
has perfect fluid behaviour, i.e.
$$
\bar{k}_\nu=0\,,~\bar{q}_\nu=0\,,~\bar{\pi}_{\nu\kappa}=0
$$
and by (\ref{bprop2}), there is no particle flux, i.e. $k_\nu=0$.

The BGK multipole equations (\ref{bhe1}) and (\ref{bhe2}) reduce to 
\begin{eqnarray}
2e^{-\Gamma}{\partial\over\partial E}\left(\lambda^3
\sigma ^{\nu\kappa}H_{\nu\kappa}\right)
+5\lambda^3\Theta{\partial \bar{F}\over\partial E}
-15\lambda E {\partial \bar{F}
\over\partial t}
&=&0\label{bhe1'}\\
{\partial\over\partial E}\left(\lambda^4\sigma ^{\kappa\rho}
F_{\nu\kappa\rho}\right)&=&0\label{bhe2'} 
\end{eqnarray} 
The higher order BGK multipole equations 
become conditions on the fourth and higher
multipoles which may always be satisfied,  since these 
multipoles are otherwise unrestricted. 
The condition (\ref{bhe2'}) gives 
$$
\sigma ^{\kappa\rho}F_{\nu\kappa\rho}=0
$$ 
while (\ref{bhe1'}) is a constraint on $\bar{F}$ 
and $V_\nu$.  Once these are specified subject to (\ref{bhe1'}), 
the remaining field
equations determine $g_{\nu \nu }(t)$ in principle, 
thus completing the Einstein--BGK solution. Using the fact that
$$
\sigma^\nu{}_\kappa=\mbox{diag}\left({\dot{X}\over X}-{1\over3}\Theta,
{\dot{Y}\over Y}-{1\over3}\Theta,{\dot{Z}\over Z}-{1\over3}\Theta
\right)
$$
(\ref{bhe1'}) leads to the new result
\begin{eqnarray}
&&2e^{-\Gamma}\left({\dot{X}\over X}-{\dot{Z}\over Z}\right)
{\partial\over\partial E}\left(\lambda^3V_1\right)+2e^{-\Gamma}
\left({\dot{Y}\over Y}-{\dot{Z}\over Z}\right){\partial\over
\partial E}\left(\lambda^3 V_2\right) \nonumber\\
&&{}+5\left({\dot{X}\over X}+{\dot{Y}\over Y}+{\dot{Z}\over Z}\right)
\lambda^3{\partial \bar{F}\over\partial E}-15\lambda E
{\partial\bar{F}\over\partial t}=0
\label{big}\end{eqnarray}
It is clearly possible to find $\bar{F},V_\nu$ that satisfy the
single linear equation (\ref{big}).

\section{Anisotropy generation in FRW cosmologies}

Using the FRW and Bianchi I results we now construct a model in which 
an Einstein--Boltzmann
solution in FRW spacetime evolves into an Einstein--Boltzmann solution 
in Bianchi I spacetime.  It
is assumed that the universe initially has FRW geometry with a 
matter distribution that is
described by an anisotropic distribution function which is also 
compatible with FRW geometry.
Ellis and Matravers \cite{em90}, \cite{em92},  \cite{me} 
propose two mechanisms whereby the
anisotropy of the matter distribution is communicated to the universe.  
The first applies to a
universe that is initially FRW with an equilibrium particle 
distribution that is inhomogeneous and
anisotropic.  The anisotropy and inhomogeneity is not communicated 
to the spacetime geometry. 
The particle distribution is effectively collision--free, 
because the particles are assumed to enjoy asymptotic
freedom under the high temperature conditions of the early universe.  
As the
universe expands,  the temperature drops and the collisions 
become significant, allowing the
anisotropy to be communicated to the geometry.  
In the second mechanism, while the 
temperature is very high
the particles effectively behave as if their rest mass is zero.  
As the universe expands and cools, 
and the threshold energies of different particles are reached,  
their rest mass becomes
significant.  Again, this effective change in the zero rest mass 
allows the inhomogeneity and
anisotropy to be communicated.  We show that the generation of 
anisotropy can be modelled
starting from an anisotropic {\em and}  
non--equilibrium BGK solution.  
A non--equilibrium model
allows us to overcome some of the drawbacks of the equilibrium model 
-- for example,  the
difficulty in motivating a collision--free early--universe phase. 

We assume that the universe initially ($t\leq t_0)$ has FRW geometry 
and that the particle
distribution is given by (\ref{rwcovh}). This distribution function is
dynamically anisotropic and has Bianchi I symmetry, 
i.e $f=f(t,p_\nu)$.  For the high 
temperature conditions in the
early universe, the particle rest mass is insignificant and 
an Einstein--BGK solution may be
chosen that is consistent with the Bianchi I solution (\ref{b1eg}).  
Furthermore, the local equilibrium $\bar{f}$ to which $f$ is close 
is a collision--dominated Planckian distribution, in keeping with
standard physics of the early universe (and not requiring asymptotic
freedom or other exotic processes).

By (\ref{econ}), (\ref{aw2}), 
(\ref{stressaw}) we can take
\begin{equation}
H=0, \mbox{  }H_\nu=0, \mbox{  }\int _0^\infty w^3H_{\nu\kappa}(w)dw=0
\label{rwb11}
\end{equation}
where $w=R(t)E$.
Thus under high temperature conditions for which $m=0$,  
it is possible to find distribution functions
with non--zero quadrupole, but for which the condition 
$\pi _{\nu\kappa}=0$ is
satisfied.  We need non--zero $H_{\nu\kappa}$ in order to 
`switch on' $\pi _{\nu\kappa}$ (the mechanism for
this will be described below).  Then by (\ref{eos}) 
the shear anisotropy $\sigma _{ij}$ will emerge
and the geometry will evolve to a Bianchi I phase.  
By (\ref{b1eg}) and (\ref{rwb11}),  we could
take $H_{\nu\kappa}$ of the form
\begin{equation}
H^\nu{}_\kappa(w)=\mbox{diag}\left[V_1(w),V_2(w),-V_1(w)
-V_2(w)\right]
\label{rwb12} 
\end{equation}
where $V_\nu$ are non--zero but satisfy 
\begin{equation}
\int _0^\infty w^3V_\nu(w)dw=0
\end{equation}
Thus for $t\leq t_0$ the universe is considered FRW,  
while for $t\geq t_0$ it is Bianchi I.  On
the 
hypersurface $t=t_0$,  the geometry and the distribution function 
symmetry must satisfy both the
conditions for FRW and Bianchi I spacetimes.  This is the case 
if the following matching
conditions are satisfied \cite{me}:
\begin{equation}
\dot{X}(t_0)=\dot{Y}(t_0)=\dot{Z}(t_0) \mbox{ and } 
X(t_0)=Y(t_0)=Z(t_0)
\label{rwb13} 
\end{equation}
As discussed in \cite{me} and \cite{em92}, these matching 
conditions always have a solution.
Then this solution becomes the initial conditions for the 
Bianchi I solution, which is governed by existence
and uniqueness theorems for the Einstein--Boltzmann equations 
(see \cite{em92}). 
The transition from
effectively massless to effectively massive behaviour will
take place over a cosmic time $\delta t$ which is much less than
the expansion time. Evolution away from
FRW is triggered as soon as the mass becomes dynamically significant,
i.e. after time $\delta t$. 
In practice we are treating
the transition as
instantaneous, in a similar way to standard models of 
electron--positron annihilation. Our simple model 
illustrates that in principle, finely tuned anisotropy in the
matter distribution can be unlocked dynamically to generate
anisotropy in the spacetime geometry.

Note that 
the evolution away from FRW occurs entirely within
the Bianchi I phase, as the shear grows from zero. 
At the transition,
$\sigma _{ij}(t_0)=0$,  and $\sigma _{ij}$ remains zero unless 
$\pi _{ij}$ becomes non--zero to
force the universe to evolve to the Bianchi I geometry.  In the 
Bianchi I phase $t\geq t_0,$ the
solution is given by (\ref{bicovh}) and it clearly reduces to the 
FRW solution at $t=t_{0 }$ as a
result of the matching conditions (\ref{rwb13}). Because $w$ 
becomes isotropic and reduces to
the FRW form with the application of the matching conditions,  
the Bianchi I equilibrium solution
$\bar f$ reduces to the FRW equilibrium solution at $t_0$.

In summary, the mathematical model is the following.
The collision model is AW:  
\begin{equation}
\Gamma (t)=\int ^t_0{dt'\over\tau (t')} \mbox{   for all  } t\geq 0
\label{rwb14} 
\end{equation}
The particle energy $E=-u_ip^i=p^0$ is
\begin{equation}
E=\left\{
\begin{array}{ll}
R^{-1}(t)\left[(p_1)^2+(p_2)^2+(p_3)^2\right]^{1/2} 
\mbox{   }(m=0) &t\leq t_0 \\ 
{}&{}\\
\left[m^2+X^{-2}(t)(p_1)^2+Y^{-2}(t)(p_2)^2+
Z^{-2}(t)(p_3)^2\right]^{1/2} &t\geq t_0
\end{array}
\right.
\end{equation}
with $w^2=(p_1)^2+(p_2)^2+(p_3)^2$ for all $t\geq 0$. 
(Note that $w=R(t)E$ for $t\leq t_0$.)

The particle distribution function is given by
$$
 f(t,p_\nu )=\bar{f}(w)+h(p_\nu )e^{-\Gamma (t)}
$$
where
\begin{equation}
\bar{ f}=\left\{
\begin{array}{ll}
\bar{F}(R(t)E) &t\leq t_0\\ 
\bar{F}(t,E)+\bar{F}_\nu(t,E)e^\nu+\cdots &t\geq t_0
\end{array}
\right.
\end{equation}
with $\bar{F}_{\nu \cdots }(t_0,E)=0$, 
$\bar{F}(t_0,E)=\bar{F}(R(t_0)E)$, and
$$
h(p_\nu )=H(t,E)+H_\nu(t,E)e^\nu+\cdots 
$$
with $H_{\nu\cdots}(t,E)=H_{\nu\cdots}(R(t)E))$
for $t\leq t_0$.  Note that
in the FRW phase ($t\leq t_0$) each 
$\bar{F}_{\nu\cdots}$ and $H_{\nu\cdots}$ is a Liouville 
solution (hence functions of $w$ only),  because the multipoles
decouple in $L[f]=0$ due to $\sigma _{ij}=0$.  This is no longer true 
in the Bianchi I phase ($t\geq t_0)$ and 
hence $\bar{F}_{\nu\cdots}$ and $H_{\nu\cdots}$ 
are functions of $t$ and
$E$. 

The distribution is specified to yield an 
Einstein--BGK solution for $t\geq 0$:
$$
H(t,E)=0=H_\nu(t,E)=0 \quad
\bar{F}_\nu(t,E)=0=\bar{F}_{\nu\kappa}(t,E)
$$
The resultant solution is
\begin{equation}
f(t,p_\nu )=\bar{F}(t,E)+e^{-\Gamma (t)}H_{\kappa\rho}(t,E)e^\kappa
e^\rho+\cdots 
\label{rwb15} 
\end{equation}
with $\bar{F}=\bar{F}(R(t)E)$, $H_{\nu\kappa}=H_{\nu\kappa}(R(t)E)$ 
for $t \leq t_0$, and $H_{\nu\kappa}$ obeys
$$
H^\nu{}_\kappa(t,E)=\mbox{diag}\left[V_1(t,E),V_2(t,E),
-V_1(t,E)-V_2(t,E)\right] 
$$
where for $t\leq t_0$:
\begin{eqnarray}
V_\nu (t,E)&=&V_\nu (R(t)E) \label{rwb16} \\
\int _0^\infty w^3V_\nu(w)dw&=&0 \nonumber
\end{eqnarray}
and for $t\geq t_0$:
\begin{eqnarray}
\pi^\nu{}_\nu(t)&=&{{8\pi }\over{15}}e^{-\Gamma (t)}\int _m^\infty
(E^2-m^2)^{3/2}V_\nu(t,E)dE  \mbox{  (no sum)}
\label{rwb17} \\
&\neq &0 \nonumber
\end{eqnarray}
Finally, $\bar{F}$ and $V_\nu$ are subject for $t\geq t_0$ to
the constraint (\ref{big}):
\begin{eqnarray}
&&2e^{-\Gamma}\left({\dot{X}\over X}-{\dot{Z}\over Z}\right)
{\partial\over\partial E}\left(\lambda^3V_1\right)+2e^{-\Gamma}
\left({\dot{Y}\over Y}-{\dot{Z}\over Z}\right){\partial\over
\partial E}\left(\lambda^3 V_2\right) \nonumber\\
&&{}+5\left({\dot{X}\over X}+{\dot{Y}\over Y}+{\dot{Z}\over Z}\right)
\lambda^3{\partial \bar{F}\over\partial E}-15\lambda E
{\partial\bar{F}\over\partial t}=0
\label{big'}\end{eqnarray}

It is clearly possible to find $\bar{F}$, $V_\nu$ such 
that (\ref{rwb16}) -- (\ref{big'}) are satisfied (and
therefore all field and conservation equations will be satisfied).  
The solution has zero number
flux, bulk viscosity and energy flux for all $t\geq 0$. 

The early universe  
temperature is high enough that the
particles 
have $m=0$, and the matter distribution is 
described by (\ref{rwb15}), consistent with
an FRW spacetime geometry.
The conditions (\ref{big}), (\ref{rwb16}) and (\ref{rwb17}) can be
satisfied by appropriate $H_{\nu \kappa}\not= 0$ and thus the 
field and conservation conditions are
satisfied 
for both the FRW and Bianchi I phases.  As the universe expands and 
cools below the threshold
energy (at $t=t_0$) for the particles under consideration, the 
distribution is no longer effectively of 
zero rest mass particles and the condition (\ref{rwb16}) is no 
longer satisfied.  Condition
(\ref{0stress}) is required to ensure vanishing anisotropic stress 
in the FRW phase (i.e. the field
equation $\pi _{\nu \kappa}=0$ is satisfied).  
For the AW collision model (\ref{rwb16}) is satisfied by
$H_{\nu \kappa}\not= 0$ only if $m=0$, by (\ref{0stress}).  
Thus,  as soon as the particle rest mass is no
longer effectively zero,  the anisotropic stress becomes non--zero 
($\pi _{\nu \kappa}\not= 0$) and is given
by (\ref{rwb17}).  As a result shear anisotropy emerges,  
forcing the universe to evolve away from
the FRW to the Bianchi I phase.  

During the phase $t\leq t_0$ the collision rate is high and 
therefore $e^{-\Gamma }$ may become
small.  This forces
the non-equilibrium, anisotropic distribution function 
$f(t,p_\nu )$ to approach the isotropic,
collision--dominated equilibrium distribution $\bar{f}(w)$.  
However, as long as $e^{-\Gamma }>0$
(even if it is very small) the model presented here works.  
The model therefore represents a high
temperature situation where the matter distribution is nearly 
isotropic (forced by the high collision
rate).  The distribution function has Bianchi I symmetry but 
satisfies all the conditions for an FRW
universe.  As the universe cools the particle threshold energy 
is reached and the FRW condition
is no longer satisfied. The remnant of the initial anisotropy 
therefore acts as the seed for the
change to Bianchi I geometry when the particle mass becomes 
significant,  communicating the
anisotropy of the distribution function to the spacetime geometry.  
Hence,  this model suggests a
mechanism by which anisotropy present in the radiation era could act 
as the seed for anisotropy
generation in the spacetime geometry.
The small anisotropy in the
microwave background radiation could be a physical example of 
the remnant anisotropy
considered here (see \cite{cf} for a related discussion).

\ack{
FPW thanks Portsmouth University and
RM thanks Natal University, for funding and hospitality.
RM was supported by a Portsmouth University research grant.}

\end{document}